\documentclass[twocolumn]{webofc}
\usepackage[varg]{txfonts}   

\usepackage{graphicx}
\newcommand{\nn}{\nonumber}
\newcommand{\be}{\begin{equation}}
\newcommand{\ee}{\end{equation}}
\newcommand{\ba}{\begin{eqnarray}}
\newcommand{\ea}{\end{eqnarray}}
\newcommand{\ci}[1]{\cite{#1}}
\def\gev{\,{\rm GeV}}
\def\mev{\,{\rm MeV}}
\newcommand{\tw}{\textwidth}
\newcommand{\req}[1]{(\ref{#1})}
\def\={\,=\,}
\newcommand{\lsim}{\raisebox{-4pt}{$\,\stackrel{\textstyle <}{\sim}\,$}}
\newcommand{\gsim}{\raisebox{-4pt}{$\,\stackrel{\textstyle
      >}{\sim}\,$}}

\woctitle{Transversity 2014}
\begin{document}
\title{Status of DVMP, DVCS and GPDs }
%
%

\author{P.~Kroll \inst{1}\fnsep\thanks{\email{kroll@physik.uni-wuppertal.de}}} 

\institute{Fachbereich Physik, Universit\"at Wuppertal, 42097 Wuppertal,
Germany and Institute f\"ur Theoretische Physik, Universit\"at Regensburg, 93040
Regensburg, Germany}

\abstract{%
The analysis of exclusive meson leptoproduction (DVMP) within the handbag approach is 
reviewed and the parametrization of the generalized parton distributions (GPDs) is
discussed in some detail with the main interest focused of the GPDs $H$ and $E$.
Applications of the GPDs extracted from DVMP to other hard exclusive processes as for
instance deeply virtual Compton scattering (DVCS) and an evaluation of Ji's sum rule 
are also presented.  }
\maketitle
\section{Introduction}
\label{intro}
\begin{figure}[t]
\begin{center}
\includegraphics[width=0.3\tw]{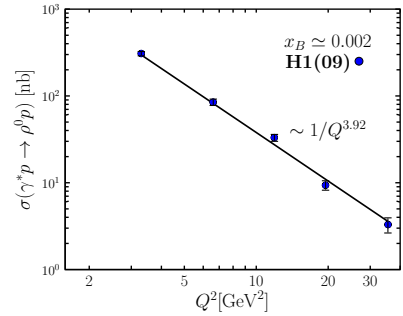}
\caption{The cross section for $\rho^0$ electroproduction versus $Q^2$ at 
$x_B \simeq 0.002$. Data are taken from \ci{h109} and compared to a power-law fit.}
\label{fig:1}
\end{center}
\end{figure}

The handbag approach to hard exclusive leptoproduction of photons and mesons off 
protons has extensively been studied during the last fifteen years. It turned out 
that the handbag approach allows for a detailed analysis of cross sections, 
asymmetries and spin density matrix elements (SDME) for these processes. The 
handbag approach is based on factorization of the process amplitudes in a hard 
subprocess, e.g.\ $\gamma^*q\to \gamma (M) q$, and soft hadronic matrix elements 
parametrized in terms of GPDs. This factorization property has been shown to hold 
rigorously in the collinear limit for large photon virtuality, $Q$, and large 
energy, $W$, but fixed Bjorken-$x$, $x_B$ \ci{collins-freund,collins-FS}. However,
power corrections to these asymptotic results are not under control. It is therefore
unclear at which values of $Q^2$ and $W$ the asymptotic results apply. In fact, 
there are strong effects in meson leptoproduction which are not in accord with the 
asymptotic predictions. Thus, for instance, the contribution from longitudinally 
polarized virtual photons to likewise polarized vector (or pseudosalar) mesons 
transitions ($\gamma^*_L\to V^{\phantom{*}}_L(P)$) dominate asymptotically; the ratio 
of the longitudinal and transverse cross sections ($R=\sigma_L/\sigma_T$) grows 
proportionally to $Q^2$. Experimentally \ci{h109}, $R$ for $\rho^0$ production only 
amounts to about 2 for $Q^{\,2}\lsim 10\,\gev^2$, i.e.\ contributions from 
transversely polarized photons are not small. For $\omega$ production transverse 
photons even dominate \ci{hermes-omega},  $R(\omega)$ is only about 0.3 for 
$2\,\gev^2\lsim Q^2 \lsim 4\,\gev^2$. For $\pi^0$ production transverse photons 
probably dominate as well \ci{GK6}. The amplitudes for $\gamma^*_L\to \rho^0_L$ 
transitions do also not plainly agree with the asymptotic picture which predicts 
the scaling law $\sigma_L \propto 1/Q^{\,6}$ (modulo powers of $\ln{Q^{\,2}}$ from 
evolution and the running of $\alpha_s$) at fixed Bjorken-$x$. As can be seen from 
Fig.\ \ref{fig:1} the data~\footnote{
Since $R$ is slightly increasing with $Q^{\,2}$ $\sigma_L$ is even flatter than 
$1/Q^{\,4}$.}
for the $\rho^0$ cross section \ci{h109} rather fall as $\lsim 1/Q^{\,4}$. Another 
example of corrections to the asymptotic results for the $\gamma^*_L\to V_L(P)$ 
amplitudes is set by the strong contributions from the pion pole to $\pi^+$ production
that has been observed experimentally \ci{hermes-pip,F2}.  
In this talk I am going to report on an extraction of the GPDs from DVMP  
\ci{GK6,GK2,GK3,GK5}. In this analysis the GPDs are constructed from double 
distributions (DDs)\ci{mueller94, rad98} and the partonic subprocesses are computed 
within the modified perturbative approach in which quark transverse degrees of 
freedom as well as Sudakov suppression \ci{li} are taken into account in order to 
model power corrections. As explained above these corrections are needed for 
instance in order to change the asymptotic $1/Q^{\,6}$ fall of the longitudinal 
$\rho^0$ cross section in an effective $1/Q^{\,4}$ behavior. The emission and 
re-absorption of the partons by the protons are treated collinearly to the proton 
momenta in \ci{GK6,GK2,GK3,GK5}. From the analyses of the longitudinal cross sections 
for $\rho^0$ and $\phi$ production the GPD $H$ has been extracted \ci{GK2}. The 
transverse target spin asymmetries for $\rho^0$ production provide information on 
the GPD $E$. Generalizations of the handbag approach to $\gamma^*_T\to V^{\phantom{*}}_T$ 
and $\gamma^*_T\to V^{\phantom{*}}_L(P)$ transitions allow for a study of further GPDs 
($\widetilde{H}, \widetilde{E}, H_T, \bar{E}_T$). The extracted set of GPDs are 
subsequently be applied to calculate other hard exclusive processes free of adjustable 
parameters. An example of such an application is DVCS. Most of the observables of this 
process are under control of the GPD $H$. Nevertheless, the transverse target spin 
asymmetries in DVCS provide an additional constraint on $E$. The extracted GPDs $H$ 
and $E$ allow for an evaluation of the angular momenta the partons inside the proton 
carry. In the following sections these analyses and studies are described in some detail.
\section{The double distribution representation}
\label{sec:1}
There is an integral representation of the GPDs in terms of DDs \ci{mueller94,rad98}:
\ba
\hspace*{-0.05\tw}K^{\,i}(x,\xi,t)&=&\int_{-1}^{1}\, d\rho\,\int_{-1+|\rho|}^{1-|\rho|}\, 
        d\eta \,\delta(\rho+\xi\eta -x)\, r_i(\rho,\eta,t)  \nn\\
            &+& D_i(x,t)\, \Theta(\xi^2-x^2)
\label{eq:int-rep}
\ea
where $K^{\,i}$ is some GPD. According to Diehl and Ivanov \ci{diehl-ivanov} there is an 
additional factor $x/\rho$ in \req{eq:int-rep} for $\widetilde{H}^g$ and $\widetilde{E}^g$. 
The last term in \req{eq:int-rep} is the so-called $D$-term \ci{pol99} which appears for 
the GPDs $H$ and $E$. As a consequence of time-reversal invariance the gluonic $D$-term 
is an even function of $x$ and the quark one an odd function. The advantage of the DD 
representation is that polynomiality of the GPDs is automatically satisfied. 

A frequently used ansatz for the DD, $r_i$, associated with $K^i(x,\xi,t)$ is \ci{mus-rad}
\be
r_i(\rho,\eta,t)\=K^{\,i}(\rho,\xi=0,t) w_i(\rho,\eta)\,.
\ee
The weight function, $w_i$, that generates the skewness dependence of the
GPD, is assumed to be 
\be
w_i(\rho,\eta)\= \, \frac{\Gamma (2n_i+2)}{2^{2n_i+1}\Gamma^2 (n_i+1)} \,
\frac{[(1-|\rho |)^2-\eta^2]^{n_i}}{(1-|\rho |)^{2n_i+1}} 
\label{eq:weight}
\ee   
(in \ci{GK2,GK3,GK5}: $n=1$ for valence quarks and 2 for sea quarks and gluons).
The zero-skewness GPD for $\rho\geq 0$ is parametrized as its forward limit, 
$K^{\,i}(\rho,\xi=t=0)=k^{\,i}(\rho)$, multiplied by an exponential in Mandelstam $t$
\be
K^{\,i}(\rho,\xi=0,t) \= k^{\,i}(\rho)\, \exp \big[ t f_{i}(\rho) \big]
\label{eq:zero-skewness}
\ee
and is to be suitably continued to negative $\rho$. For $H$, $\widetilde H$ and 
the transversity GPD $H_T$ the forward limits are the corresponding unpolarized, 
polarized and transversity parton distributions (PDFs),  
respectively~\footnote{
Note that by definition the forward limits of the gluonic GPDs have an extra factor
of $x$, e.g. $H^g(x,\xi=t=0)=xg(x)$.}. 
The forward limits of the other GPDs ($E$, $\widetilde E$, $\bar{E}_T$) which are not 
accessible in deep inelastic lepton-nuleon scattering (DIS), are parametrized in a 
fashion analogously to the PDFs
\be
k^{\,i}(\rho)\=N_i\, \rho^{-\delta_i}\,(1-\rho)^{\beta_i}
\label{eq:forward}
\ee
with free parameters $N_i$, $\delta_i$ and $\beta_i$ to be adjusted to data on
exclusive reactions. In order to perform the DD integral \req{eq:int-rep} 
analytically the PDFs are expanded (for $H$, $\widetilde H$, $H_T$): 
\be
 k^i(\rho)\=\rho^{-\delta_i}(1-\rho)^{2n_i+1} \sum_{j=0}^3 c_{ij}\,\rho^{j/2}\,.
\label{eq:expansion}
\ee 
For quarks $\delta_i$ equals a Regge-like intercept $\alpha_i$ 
while, for gluons, $\delta_g=\alpha_g-1$ where $\alpha_g$ is a Pomeron-like 
intercept. 
 
The profile function, $f_{i}(\rho)$, is parametrized in a Regge-like manner
\be
f_{i}(\rho) \= -\alpha_{i}^\prime \ln{\rho} + B_i
\label{eq:profile}
\ee
where $\alpha_i^\prime$ can be regarded as the slope of an appropriate Regge trajectory
and $B$ parametrizes the $t$ dependence of its residue. This profile 
function is a simplified version of a more complicated one that has been proposed in
\ci{DFJK4,DK13} 
\be
f_i(\rho) \= \big(-\alpha_i^\prime \ln{\rho} + B_i\big)\,(1-\rho)^3 
             + A_i\,\rho(1-\rho)^2\,.
\label{eq:profile-dfjk4}
\ee

\begin{figure}[t]
\begin{center}
\includegraphics[width=0.3\tw]{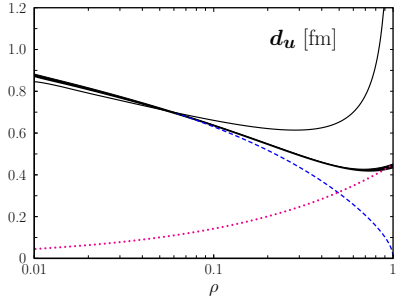}
\caption{The distance $d_u$ evaluated from the profile functions \req{eq:profile} 
(dashed line) and \req{eq:profile-dfjk4} (solid line). The separate contributions 
from the low-$\rho$ (dashed) and the large-$\rho$ (dotted) term in the profile 
function \req{eq:profile-dfjk4} are also shown. The figure is taken from \ci{DK13}.}
\label{fig:2}
\end{center}
\end{figure}

In order to elucidate the physics underlying the ans\"atze for the profile function
let us consider the Fourier transform of the GPD $H^q(\rho,\xi=0,t)$ with respect to 
the momentum transfer ${\bf \Delta}_\perp$
\be
q(\rho,{\bf b}_\perp)\=\int \frac{d^2{\bf \Delta}_\perp}{(2\pi)^2} 
                  e^{-i{\bf b}_\perp{\bf \Delta}_\perp} 
                   H^q(\rho,\xi=0,t=-\Delta^2_\perp)\,.
\ee
According to Burkardt \ci{burkhardt,burkhardt02}, $q(\rho,{\bf b}_\perp)$ possess a 
density interpretation. The variable ${\bf b}_\perp$ is the transverse distance 
between the struck quark and the hadron's center of momentum defined by 
$\sum \rho_j{\bf b}_{\perp j}=0$. Evidently, quarks with a large momentum fraction 
$\rho_j$ must have a small transverse distance in that frame. In other words there 
is a correlation in $q(\rho,{\bf b}_\perp)$ between $\rho$ and $b_\perp$. In the limit 
$\rho\to 1$ $H^q$ becomes $t$ independent~\footnote{
The profile function \req{eq:profile} don't possess this property except $B_i=0$.}.
An estimate of the proton's transverse radius is provided by the average distance 
between the struck quark and the cluster of spectator partons:
\be
d_q(\rho) \= \frac{\sqrt{\langle b^2_\perp \rangle^q_\rho}}{1-\rho}\,.
\ee
For the ansatz \req{eq:zero-skewness} the average distance reads
\be
d_q(\rho)\=2\frac{\sqrt{f_q(\rho)}}{1-\rho}\,.
\ee
We see that the profile function \req{eq:profile} is singular for $\rho\to 1$ while 
\req{eq:profile-dfjk4} leads to $d_q\to 2\sqrt{A_q}$ in this limit. The average 
distances for $u$ quarks obtained from both these profile functions are shown in 
Fig.\ \ref{fig:2}. The Regge-like profile function is a reasonable approximation to 
\req{eq:profile-dfjk4} at small $\rho$. 

At zero skewness  $H^q$  exhibits a strong correlation between $\rho$ and $t$; the large 
$-t$ behavior of $H^q$ is under control of large $\rho$. The reason for this $\rho-t$ 
correlation in the parametrization of the GPDs described above, is easily understood. At 
small $\rho$ the PDF behaves $\sim \rho^{-\delta_q}$ with $\delta_q\simeq 0.5$ for valence 
quarks. For a simple $\rho - t$ factorized ansatz as has been used at the beginning of 
the handbag physics \ci{VGG,freund}
\be
H^q(\rho,\xi=0,t) \sim q(\rho) F^q(t)
\label{eq:vgg}
\ee
the GPD possess that $\rho^{-\delta_q}$ singularity at all $t$. For the ans\"atze 
\req{eq:profile} and \req{eq:profile-dfjk4} the small $\rho$ behavior of the GPD is 
changed in
\be
H^q(\rho,\xi=0,t) \sim \rho^{-(\delta_q+\alpha_q' t)}\,.
\label{eq:fac}
\ee
The $\rho^{-\delta_q}$ singularity occurring at $t=0$ becomes milder with increasing $-t$ 
and turns finally in a zero for $-t$ larger than $\delta_q/\alpha_q'$. Given that for 
$-t\gsim \delta_q/\alpha_q'$ the GPD, parametrized as in \req{eq:zero-skewness}, only 
possesses nodes at the end points, it exhibits a pronounced maximum at a position that 
shifts towards higher $\rho$ with increasing $-t$. This property of the zero-skewness
GPDs is transfered to the case of $\xi\neq 0$ through \req{eq:int-rep} although to a
lesser degree. For all $\xi>0$ the GPD is peaked at a position $<\xi$ which increases 
with increasing $-t$. The peak becomes less pronounce with increasing $\xi$ (at fixed $t$)
and with increasing $-t$ (at fixed $\xi$). Because of the $\rho - t$ correlation the 
profile function \req{eq:profile} can only be applied at small $-t$; it is 
\textit{unphysical} at large $-t$. 

In an  alternative parametrization the GPDs are decomposed in terms of $t$-channel
partial wave amplitudes. Each partial wave is parametrized in a Regge-like manner
\ci{MM,kmls}.

\section{Extraction of the GPD $H$ from DVMP}
\label{sec:2}
\begin{figure}[t]
\centering
\includegraphics[width=0.41\tw]{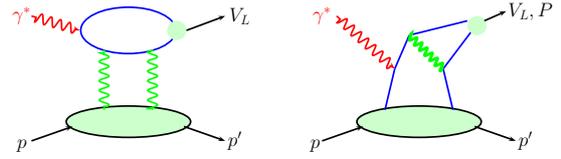}
\caption{Typical leading-order Feynman graphs for exclusive meson production.}
\label{fig:0}
\end{figure}

The asymptotically dominant $\gamma_L^*\to V^{\phantom{*}}_L$ amplitudes~\footnote
{The light-cone helicities are labeled by their signs or by zero.}
 read ($V=\rho^0, \omega, \phi$,
the generalization to other vector mesons is straightforward)
\ba
\hspace*{-0.04\tw} {\cal M}^{\,V}_{0+,0+}\hspace*{-0.01\tw} &=&\hspace*{-0.01\tw} \frac{e_0}2 
      \sqrt{1-\xi^2}\sum_{q=u,d,s}e_q {\cal C}_V^q
\Big[\langle H^g_{\rm eff} \rangle_V   + \langle H^q_{\rm eff}\rangle_V\Big]\,,\nn\\
\hspace*{-0.04\tw} {\cal M}^{\,V}_{0-,0+}\hspace*{-0.01\tw} &=&\hspace*{-0.01\tw} -\frac{e_0}2
      \frac{\sqrt{t_0-t}}{2m} \sum_{q=u,d,s} e_q{\cal C}_V^q
\Big[\langle E^g \rangle_V + \langle E^q\rangle_V\Big] 
\label{eq:amplitudes}
\ea
where $e_q$ denote the quark charges in units of the positron charge, $e_0$,
and $m$ the mass of the nucleon. Because of parity conservation it suffices to consider 
only the amplitudes with helicity $1/2$ of the initial state proton. The non-zero flavor 
weight factors read
\be
{\cal C}^u_{\rho^0}=-{\cal C}^u_{\rho^0}={\cal C}^u_{\omega}={\cal C}^d_{\omega}=1/\sqrt{2}\,,
\qquad {\cal C}^s_{\phi}=1\,.
\ee
The GPD $H_{\rm eff}$ for quarks and gluons represents the combination
\be
H_{\rm eff}=H-\frac{\xi^2}{1-\xi^2}E\,.
\ee
In \ci{GK2} a meson-mass correction is taken into account in the relation between skewness 
and $x_B$
\be
\xi\=\frac{x_B}{2-x_B}\big[1+m_V^2/Q^2\big]\,.
\label{eq:xi-xb}
\ee
The minimal value of $t$ allowed in the process is
\be
t_0\=-4m^2\frac{\xi^2}{1-\xi^2}\,.
\ee
The convolutions $\langle K\rangle$ in \req{eq:amplitudes} read
\be
\langle K^i \rangle_V\=\sum_\lambda \int_{x_i}^1 dx\, 
   {\cal H}^{Vi}_{0\lambda,0\lambda}(x,\xi,Q^2,t=0) K^i(x,\xi,t)
\label{eq:convolution-mesons}
\ee
where $i=q,g$ and $x_q=-1$, $x_g=0$. The last item to be specified is the subprocess 
amplitude ${\cal H}$ for partonic helicity $\lambda$. In \ci{GK2} it is calculated to 
leading-order of perturbation theory (see Fig.\ \ref{fig:0}) taking into account quark 
transverse momenta, ${\bf k}_\perp$, in the subprocess and Sudakov suppressions. 
Since the latter involves a resummation of all orders of perturbation theory in 
next-to-leading-log approximation \ci{li} which can only be efficiently performed in 
the impact-parameter space canonically conjugated to the $k_\perp$-space, one is forced 
to work in the {\bf b}-space. Hence,
\ba
\hspace*{-0.03\tw}
{\cal H}^{Vi}_{0\lambda,0\lambda}&=&\int d\tau d^2b\,\hat{\Psi}_{V}(\tau,-{\bf b}) 
     \hat{F}^i_{0\lambda,0\lambda}(x,\xi,\tau,Q^2,{\bf b}) \nn\\
      &\times& \alpha_s(\mu_R) \exp{\big[-S(\tau,{\bf b},Q^2)\big]}\,.
\ea
$\hat{F}$ and $\hat{\Psi}$ are the Fourier transforms of the hard scattering kernel
and the meson's light-cone wave, respectively. For the latter quantity a Gaussian in 
${\bf b}$ is used with a parameter that describes the transverse size of the meson, 
and which is adjusted to experiment. $\mu_R$ is a suitable renormalization scale.
The modified perturbative approach utilized in \ci{GK2}, is designed in such a way 
that asymptotically the leading-twist result \ci{collins-FS} emerges. In passing I 
would like to remark that the treatment of the gluonic part of the amplitudes bears 
resemblance to the color-dipole model, see for instance \ci{martin,koepp,nikolaev}.

\begin{figure}[t]
\centering
\includegraphics[width=0.26\tw]{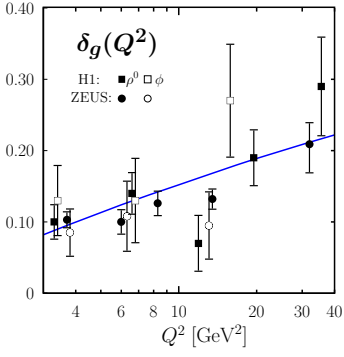}
\caption{The intercept of the gluon trajectory $\delta_g=\alpha_g-1$ versus $Q^2$. 
Data are taken from Refs.\ \ci{zeus05,chekanov07,h109}. The solid line represents the 
parametrization \req{eq:delta}.}
\label{fig:3}
\end{figure}
In \ci{GK2} the GPD $H$ is extracted from the data on the longitudinal cross section 
for $\rho^0$ and $\phi$ production in the kinematical region of \textit{small} skewness
and \textit{small} ($\xi\lsim 0.1$ and $-t\lsim 0.5\,\gev^2$). 
In this kinematical situation the contributions from the GPD $E$ to the cross section can be 
ignored. The valence-quark GPDs $H$ at zero skewness are investigated in great detail in an 
analysis of the nucleon form factors \ci{DFJK4} using the ansatz \req{eq:zero-skewness} with 
the profile function \req{eq:profile-dfjk4} and relying on the CTEQ6M(NLO) PDFs \ci{cteq6}. 
The resulting valence-quark GPDs in the small $-t$ approximation \req{eq:profile} are 
employed in \ci{GK2}. It is checked that the GPDs from the updated version of the form 
factor analysis \ci{DK13} which is based on the ABM11(NLO) PDFs \ci{abm11} and includes all
recent data of the nucleon form factors, entails only minor differences in the small 
$-t$ region.  The CTEQ6M PDFs are also utilized for the gluon and sea quarks in \ci{GK2}.
The Regge-like parameters are assumed to be same for gluons and sea quarks, 
$\alpha_g=\alpha_{\rm sea}$ and $\alpha'_g=\alpha'_{\rm sea}$, since the sea-quark and gluon 
PDFs are strongly correlated by the evolution. At very small skewness as occur in the HERA 
experiments, the cross section behaves diffractively and is dominated by the imaginary part 
of the helicity non-flip amplitude, i.e.\
\be
  \sigma_L \propto |H^g(\xi,\xi,t\simeq 0)|^2
\label{eq:cross-section}
\ee
where the mild shrinkage effect is ignored~\footnote{
With the above assumption on $\alpha_{\rm sea}$ the sea quark contribution has the same
energy dependence as the gluon one.}. 
For the DD ansatz \req{eq:int-rep} one can show that 
\be
H^g(\xi,\xi,t)\=c(\delta_g,n_g,\alpha'_gt) 2\xi g(2\xi)\,{\rm e}^{[B_g-\alpha'_g\ln(2\xi)]t}\,.
\label{eq:skewness-ratio}
\ee
Hence, at small skewness, $H^g(\xi,\xi,t\simeq 0)\sim \xi^{-\delta_g}$ with the consequence of a 
cross section obeying the power law
\be
\sigma_L \propto W^{\,4\delta_g}
\ee  
at fixed values of $Q^{\,2}$. The parameter $\delta_g$ can therefore be read off from
the HERA data \ci{zeus05,chekanov07,h109}. The results are displayed in Fig.\ \ref{fig:3} 
and compared to the fit ($Q^{\,2}_0=4\,\gev^2$)
\be
\delta_g\=0.10+0.06 \ln(Q^{\,2}/Q^{\,2}_0)-0.0027 \ln^2(Q^{\,2}/Q^{\,2}_0)\,.
\label{eq:delta}
\ee 
For the slope of the gluon trajectory $\alpha'_g$ a value of $0.15\gev^{-2}$ is taken. 
Thus, only the transverse size parameters in the wave functions for the $\rho^0$ and 
$\phi$ mesons as well as the parameter $B_g=B_{\rm sea}$ in the profile function 
\req{eq:profile} have to be fitted to experiment. For more details of the GPD 
parametrizations see \ci{GK2}. It is to be stressed that in \ci{GK2} the evolution
of the GPDs is approximated by that of the PDFs. Evolution is of importance only
at large $Q^{\,2}$ which go along with large $W$ and small $\xi$ for the available data. 
In this region the imaginary parts of gluon and sea quark contributions dominate, their 
real parts as well as the valence quark contribution are almost negligible. Because
of \req{eq:cross-section} and \req{eq:skewness-ratio} the approximate treatment of
evolution is not unreasonable. However, an update of the analysis of the longitudinal 
cross sections for $\rho^0$ and $\phi$ production should not only include more recent
sets of PDFS as for instance that of \ci{abm11} but should also make use of the full 
GPD evolution as is incorporated in the code written by Vinnikov \ci{vinnikov}.
For a detailed study of the evolution of GPDs of the type discussed in Sect.\ \ref{sec:1},
see \ci{diehl-kugler-evolution}. 

\begin{figure*}[t]
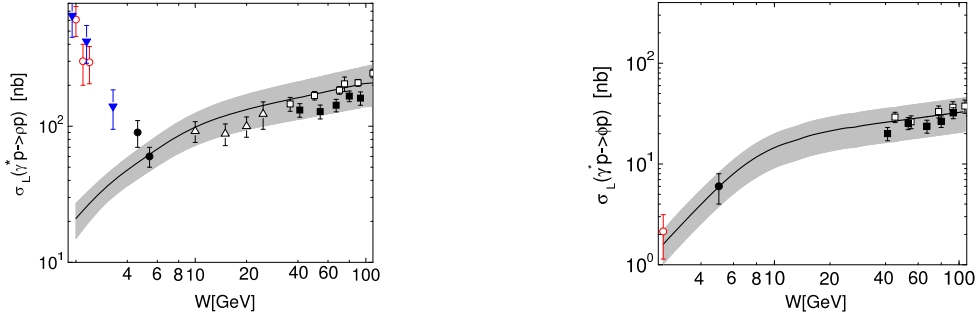

\centering
\includegraphics[width=0.29\tw]{dslrhowclas.epsi} \hspace*{0.15\tw}
\includegraphics[width=0.3\tw]{dslphiwclas.epsi}
\caption{The longitudinal cross sections for $\rho^0$ and $\phi$ production at 
$Q^{\,2}=4\,\gev^2$ and $3.8\,\gev^2$, respectively. Data are taken from 
\ci{h109,clas-rho,cornell,clas-phi}, further references can be found in \ci{GK2}. The 
solid lines with the error bands represent the results from the handbag approach.
The H1 data \ci{h109} (solid squares) are not included in the fits performed in \ci{GK2}.} 
\label{fig:4}
\end{figure*}
In Figs.\ \ref{fig:4} and \ref{fig:5} a few of the results obtained in \ci{GK2} are 
displayed. For $W>4\,\gev$ fair agreement between theory and experiment is to be seen 
for the longitudinal cross section of $\rho^0$ and $\phi$ production, integrated  
on $t$ from $t_0$ to $t_0-0.5\,\gev^2$. The error bands assigned to the theoretical results 
follow from the Hessian errors of the CTEQ6 PDFs. Exploiting other sets of PDFs (for 
instance \ci{MRST04,alekin}) one finds results which lie within the quoted error bands 
provided these PDFs are also fitted to the expansion \req{eq:expansion} with the 
experimental value \req{eq:delta} of the power $\delta_g(Q^{\,2})$. In contrast to 
$\phi$ production the handbag approach fails for $W$ below $\simeq 4\,\gev$ in the case 
of $\rho^0$ production. The strong increase of the data \ci{clas-rho,cornell} towards 
smaller $W$ is not reproduced. The kinematical region of $2\,\gev\leq W \leq 4\,\gev$ 
and $Q^{\,2}\simeq 4\,\gev^2$ is characterized by large skewness and large $-t_0$ (e.g.\
at $W=2\,\gev$: $\xi=0.45$ and $t_0=-0.89\,\gev^2$). Thus, the use of the GPDs in that 
region requires an extrapolation from the region of $\xi\lsim 0.1$ and 
$-t\lsim 0.5\,\gev^2$ where the GPDs have been fixed, to the region of large $\xi$ and 
rather large $-t$. The behavior of the handbag results for the integrated 
cross sections at low $W$  reflects the decrease of the GPDs with $-t$, 
see \req{eq:zero-skewness}. The dynamical origin of the experimentally observed behavior 
of the longitudinal cross section for $\rho^0$ production at low $W$ is unknown as yet. 
It has been conjectured in \ci{teryaev} that the $D$-term in \req{eq:int-rep} which has 
been neglected in \ci{GK2}, is responsible for it. However, it seems that this interpretation 
requires a large, nearly $t$-independent $D$-term. Both these properties seem to be in 
conflict with the findings in \ci{pasquini}. The cross sections for $\omega$ and $\rho^+$ 
production behave similar to the $\rho^0$ cross section at low $W$. It is unclear whether 
all these processes can be described by the $D$-term in a consistent way.  

In Fig.\ \ref{fig:5} the cross section for $\rho^0$ production is shown versus $Q^{\,2}$
at large $W$. In correspondence with Fig.\ \ref{fig:1} the unseparated cross section is
displayed. Good agreement with experiment is seen for $Q^{\,2}$ ranging from about 4 to
$100\,\gev^2$. The leading-twist result shown for comparison, deviates substantially from 
experiment at lower values of $Q^2$ but is close at $Q^2\simeq 100\,\gev^2$. This 
feature has already been discussed in the context of Fig.\ \ref{fig:1}. In the 
modified perturbative approach utilized in the computation of the subprocess amplitude 
\ci{GK2}, the Sudakov factor and the meson wave function generate series of power 
corrections of the type $(\Lambda_{\rm QCD}/Q)^{2n}$ and $(k_\perp/Q)^{2n}$, respectively. 
These power corrections reduce the leading-twist behavior of $\sigma$ from 
$\sim 1/Q^{\,6}$ to an effective $1/Q^{\,4}$ one which is in agreement with experiment. 
An alternative concept is advocated for in \ci{MM}. Their GPD $H$, fitted to the HERA 
data on DVMP and DVCS in collinear approximation, exhibit strong evolution effects. 
I.e.\ the reduction from the $1/Q^{\,6}$ fall to an effective $1/Q^{\,4}$ one is realized 
by powers of $\ln{Q^{\,2}}$. It remains to be seen whether this concept can be extended 
to smaller $W$. It should be mentioned that in collinear approximation a fit to only the 
DVCS data and a fit to both DVCS and DVMP data lead to different GPDs \ci{MM}.

\section{Generalizations and applications}
\label{sec:3}
\begin{figure}[t]
\centering
\includegraphics[width=0.33\tw]{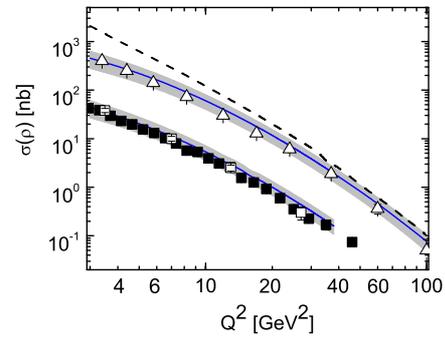}
\caption{The integrated cross section for $\rho^0$ production versus $Q^{\,2}$ at $W=75$ 
(divided by 10 for the ease of legibility)  and $90\,\gev$. For comparison the 
leading-twist result at $90\,\gev$ is also shown as a dashed line. The figure is taken 
from \ci{GK3}. For further notations see Fig.\ \ref{fig:4}.}
\label{fig:5}
\end{figure}

In \ci{GK3} the handbag approach has been generalized to the amplitudes for 
$\gamma^*_T\to V^{\phantom{*}}_T$ transitions. In collinear approximation the subprocess 
amplitudes for such transitions are infrared singular. The quark transverse momenta which 
are taken into account in the modified perturbative approach, regularize these infrared 
singularities although in a model-dependent way. The transverse amplitudes are 
suppressed by $\langle k_\perp^2\rangle^{1/2}/Q$ with respect to the longitudinal ones. 
With the $\gamma^*_T\to V^{\phantom{*}}_T$ amplitudes at disposal the transverse cross 
sections as well as some of the SDMEs for $\rho^0$ and $\phi$ leptoproduction can be 
computed. Reasonable agreement of the $\gamma^*_T\to V^{\phantom{*}}_T$ amplitudes 
with experiment is found with the exception of the relative phase between the 
longitudinal and transverse amplitudes which appears to be somewhat small.
An example of these results is shown in Fig.\ \ref{fig:5}.

The handbag approach has also been generalized to the $\gamma_T^*\to V^{\phantom{*}}_L,P$ 
transition amplitudes \ci{GK6,GK5}. The asymmetries measured with a transversely polarized 
target by the HERMES collaboration \ci{hermes-aut} for $\pi^+$ electroproduction signal the 
importance of the helicity non-flip amplitude ${\cal M}_{0-,++}$ in that process. This 
amplitude is modeled by a convolution of the transversity GPD $H_T$ and the quark 
helicity-flip subprocess amplitude which necessitates the use of a 
twist-3 meson wave function \ci{braun,beneke}:
\be
{\cal M}^{\,\pi^+}_{0-,++}\=e_0\sqrt{1-\xi^2} \int_{-1}^1 dx {\cal H}^{\,\pi^+}_{0-.++}
                       \big(H_T^u-H_T^d\big)\,.
\label{eq:HT}
\ee
This amplitude is parametrically suppressed by $\mu_\pi/Q$ with respect to the asymptotically 
leading $\gamma^*_L\to \pi^+$ amplitudes which look like \req{eq:amplitudes} for the appropriate
flavor combination with the replacement of $H_{\rm eff}$ by $\widetilde{H}_{\rm eff}$ and of 
$E$ by $\xi \widetilde{E}$. The parameter $\mu_\pi$ is large, $\simeq 2\,\gev$ at a scale 
of $2\,\gev$, since it is given by the pion mass, $m_\pi$, enhanced by the chiral condensate
\be
\mu_\pi\=\frac{m_\pi^2}{m_u+m_d}
\ee
by means of the divergency of the axial-vector current ($m_{u}$ and $m_{d}$ denote
current-quark masses).

A special feature of $\pi^+$ production is the pion-pole which contributes to the 
GPD $\widetilde E$ \ci{mankiewicz,goeke} 
\be
\widetilde{E}^{\rm pole}\=\Theta(|x|\leq\xi) \frac{F_P^{\rm pole}}{2\xi} 
                 \Phi_\pi\Big(\frac{x+\xi}{2\xi}\Big)
\label{eq:etilde-pole}
\ee
where $\Phi_\pi$ is the pion distribution amplitude and $F_P^{\rm pole}$ the contribution 
of the pion-pole to the pseudoscalar nucleon form factor. As already mentioned in the 
introduction the pion-pole contribution to $\pi^+$ production fails by order of 
magnitude if estimated through $\widetilde E$ since it is proportional to the square 
of the pion electromagnetic form factor in one-gluon exchange approximation. In 
\ci{GK5} the pion pole is therefore treated as a one-particle exchange leading to the 
same result for its contribution to the cross section except that full experimental 
value of the pion form factor appears which is about a factor of 2 to 3 larger than 
the perturbative result. The analysis of the HERMES data on $\pi^+$ production 
\ci{hermes-aut,hermes-pip} leads to a determination of the GPDs $\widetilde{H}$ and 
$H_T$ for valence quarks of, however, a lesser quality than $H$. The present data do 
not allow for a reliable determination of a non-pole contribution to $\widetilde{E}$.  

\begin{figure}[t]
\centering
\includegraphics[width=0.29\tw]{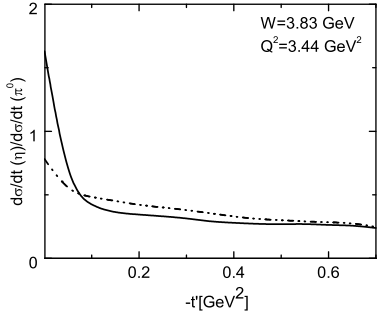}
\caption{The ratio of the $\eta$ and $\pi^0$ cross sections versus $t'=t-t_0$ for
two different parametrizations of $H_T$ (see \ci{GK6}).}
\label{fig:6}
\end{figure}
In $\pi^+$ production the transversity GPD $\bar{E}= 2 \widetilde{H}_T+E_T$  
plays a minor role while it seems to be very important in $\pi^0$ production \ci{GK6}. 
This GPD feds the amplitudes
\be
{\cal M}^{\,\pi^+}_{0+.\pm +}\=-\frac{e_0}2\frac{\sqrt{t_0-t}}{2m}\int_{-1}^{+1} dx 
                {\cal H}^{\,\pi^+}_{0-,++}\big(\bar{E}^u_T-\bar{E}^d_T\big)
\label{eq:ET}
\ee
with the same subprocess amplitude as in \req{eq:HT}. The generalization of \req{eq:HT} 
and \req{eq:ET} to other pseudoscalar mesons and even to vector mesons \ci{GK7} 
(where $\mu_V=m_V$) is straightforward. Lack of suitable data, e.g.\ the $\pi^0$
cross section at small skewness, prevents a determination of $\bar{E}_T$ from 
experiment as yet. In \ci{GK6} it is estimated by fixing its parameters with the 
help of lattice QCD results on moments of $\bar{E}_T$ \ci{qcdsf}. From this estimate  
interesting predictions for leptoproduction of pseudoscalar mesons are obtained. Thus, 
for instance, this GPD dominates $\pi^0$ production (leading for instance to 
$\sigma_L/\sigma_T\simeq 0.1$ for this process). Another example is the $\eta/\pi^0$ 
cross-section ratio which amounts to about 1/3 for $t-t_0$ not too close to zero. This 
result is in sharp contrast to a leading-twist prediction of $\gsim 1$ \ci{eides}. 
Such a large $\eta/\pi^0$ ratio may hold at the best at $t-t_0\simeq 0$ where the GPD 
$\bar{E}_T$ does not contribute (see \req{eq:ET}) and the helicity non-flip amplitudes, 
under control of $\widetilde H$ and $H_T$, take the lead. Suppose there is a dominant 
GPD. The $\eta/\pi^0$ ratio is then given by
\be
\frac{d\sigma(\eta)}{d\sigma(\pi^0)} \sim 
         \left( \frac{\langle e_uK^u +e_d  K^d\rangle_\eta}
                                  {\langle e_uK^u-e_dK^d\rangle_{\pi^0}}\right)^2\,.
\label{eq:eta-pi}
\ee
Evidently, the relative sign and magnitude of the dominant GPDs for $u$ and $d$ quarks 
determine the value of the ratio. According to the lattice result \ci{qcdsf}  
$\bar{E}_T^u$ and $\bar{E}_T^d$ have the same sign while $\widetilde{H}$ and $H_T$ 
have the opposite sign for $u$ and $d$ quarks as follows from the polarized and 
transversity PDFs \ci{dssv,anselmino}. Ignoring differences between the $\eta$ and 
the $\pi^0$ wave functions and, for a quick estimate, taking the $u/d$ ratio of the 
first moments of the zero-skewness GPD at $t=0$ as representative of the ratio of the 
convolutions, one finds the values 0.26 for a dominant $\bar{E}_T$ \ci{qcdsf}, 1.64 
for a dominant $H_T$ \ci{anselmino} and 2.40 for the case of $\widetilde{H}$ \ci{dssv} 
for the $\eta/\pi^0$ cross-section ratio. This is what one observes from Fig.\ 
\ref{fig:6}. Large skewness data from CLAS \ci{clas-pi0,clas-eta} are in agreement 
with these findings: the $\eta/\pi^0$ ratio is $\simeq 0.3$ for $-t\geq 0.1\,\gev^2$.

\begin{figure}[t]
\centering
\includegraphics[width=0.29\tw]{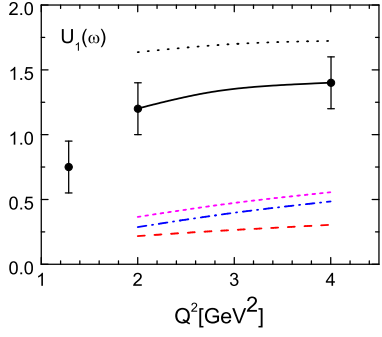}
\caption{The unnatural-parity cross section for $\omega$ leptoproduction at $W=4.8\,\gev$ and
$t-t_0=-0.08\,\gev^2$. The solid (long-dashed) line represents the handbag result for $U_1$
with (without) the pion pole. Data are taken from \ci{hermes-omega}. For other notation
see \ci{GK8}.}
\label{fig:sigmaU}
\end{figure}

\begin{table}[t]
\renewcommand{\arraystretch}{1.0} 
\caption{Status of small-skewness GPDs as extracted from meson leptoproduction data. 
At present no information is available on GPDs not appearing in the list. Except of 
$H$ for gluons and sea quarks all GPDs are only probed for scales of about 
$4\,{\rm GeV}^2$. For comparison five stars are assigned to PDFs.} 
\label{tab:1}
\begin{tabular}{| c || c | c | c |}
\hline   
GPD &  probed by &  constraints &  status \\[0.2em]
\hline
$H$(val) & {\small $\rho^0, \phi$ cross sect.} & {\small PDFs, Dirac ff} & {\small ***} 
\\[0.2em]
$H$(g,sea) & {\small $\rho^0, \phi$ cross sect.} & {\small PDFs} & {\small ***} \\[0.2em]
$E$(val)& {\small $A_{UT}(\rho^0, \phi)$} & {\small Pauli ff} & {\small **} \\[0.2em]
$E$(g,sea) &  - & \req{eq:SumRuleE} & {\small {-}}\\[0.2em]
$\widetilde{H}$ (val)  & {\small $\pi^+$ data} &{\small pol.\ PDFs, axial ff} &
{\small **} \\[0.2em]
$\widetilde{H}$(g,sea)  & $A_{LL}(\rho^0)$  & {\small polarized PDFs} & {\small
    *}\\[0.2em]
$\widetilde{E}$ (val)& {\small $\pi^+$ data} & pseudoscalar ff  
                                          & {\small *} \\[0.2em]
$H_T$(val) & {\small $\pi^+$ data} & {\small transversity PDFs} 
                                          & {\small *}\\[0.2em]
$\bar{E}_T$(val) & {\small $\pi^+$ data} & {\small -} 
                                          & {\small *}\\[0.2em]
\hline
\end{tabular} 
\renewcommand{\arraystretch}{1.0}   
\end{table}
A summary of the information about the GPDs extracted from DVMP is given in Tab.\ 
\ref{tab:1}. With these GPDs at disposal one is in the position to calculate 
observables for other hard exclusive processes exploiting the universality property 
of the GPDs. Thus, in \ci{kopeliovich} neutrino induced exclusive pion production 
has been computed. Due to the parity-violating $V-A$ structure of the electro-weak 
interactions the GPDs $H$ and $E$ also contribute to $\nu_lp\to l p\pi$. There are 
no data available at present but this process may be relevant for the MINERVA 
experiment FERMI LAB. Another example is time-like DVCS for which predictions, again
evaluated from this set of GPDs, have been given in \ci{pire13,pire14} recently. It 
would be interesting to compare them with data. In \ci{GK8} exclusive leptoproduction 
of the $\omega$ meson has been computed and compared to the SDMEs measured by HERMES 
\ci{hermes-omega}. Fair agreement is found. An important element in this calculation
is the pion-pole. As in \ci{GK5} it has been treated as an one-particle exchange; its 
calculation through \req{eq:etilde-pole} underestimates the effect as is the case for 
$\pi^+$ production. Instead of the electromagnetic form factor of the pion in the 
latter process the $\pi \omega$ transition form factor occurs now on which information 
at rather large $Q^{\,2}$ has been extracted in \ci{GK8}. The pion pole dominantly 
contributes to the $\gamma^*_T\to\omega^{*}_T$ and $\gamma^*_L\to\omega^{*}_T$ transition 
amplitudes which are suppressed by $1/Q$ and $1/Q^{\,2}$ with respect to the asymptotically 
leading $\gamma^*_L\to \omega^{*}_L$ amplitudes, respectively. As a consequence $\omega$ 
production looks very different from the asymptotic picture at $W\simeq 5\,\gev$: 
$\sigma_T>\sigma_L$ and the unnatural parity cross section is larger than the 
natural-parity one. As an example for the strength of the unnatural-parity contribution 
the ratio $U_1=d\sigma^U/d\sigma$ is shown in Fig.\ \ref{fig:sigmaU}.  Since the 
$\pi\rho^0$ transition form factor is about third of the $\pi\omega$ one \ci{chernyak84} 
the pion-pole contribution to leptoproduction of the $\rho^0$ is rather small but larger 
than what is obtained from \req{eq:etilde-pole}. It has hardly to be seen in most of the 
observables for $\rho^0$ production. Exceptions are for instance the unnatural-parity 
cross section and the relative phase between the longitudinal and transverse amplitudes 
which is enlarged from $3.1^\circ$ to $13.4^\circ$. The inclusion of the pion pole brings 
both these quantities closer to the experimental results \ci{hermes-rho}.

\section{DVCS}
\label{sec:4}
\begin{figure}[t]
\centering
\includegraphics[width=0.37\tw]{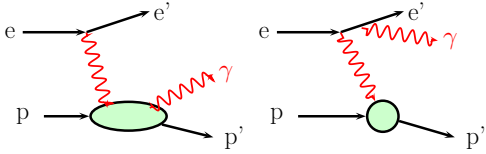}
\caption{Typical leading-order Feynman graphs for electroproduction of photons.}
\label{fig:7}
\end{figure}

Another important application of the set of GPDs extracted in \ci{GK2,GK3,GK5} is 
the evaluation of DVCS. This task has been performed by two groups: by the authors 
of \ci{kms} and by Kumericky \textit{et al.} published in \ci{eic}. Typical Feynman 
graphs for a leading-order calculation of leptoproduction of photons are shown in 
Fig.\ \ref{fig:7}. As is well-known there are two contributions to this process - 
the Bethe-Heitler (BH) contribution for which the photon is emitted from the lepton 
and the proper DVCS contribution where the photon is emitted from the proton. 
Evidently, for a collinear emission and re-absorption of quarks from the protons a 
quark transverse momentum is impossible in the subprocess $\gamma^*q \to \gamma q$. 
Therefore, the use of the collinear approximation for DVCS is consistent with the 
treatment of DVMP as described in Sect.\ \ref{sec:2}.

The square of the amplitude for leptoproduction of real photons consists of three terms
\be
| {\cal T}(l p \rightarrow l p \gamma) |^2 \= | {\cal T}({\rm{BH}})|^2 
      + {\cal T}_{\rm{I}} + | {\cal T}({\rm{DVCS}}) |^2
\label{eq:bh-interf-vcs}
\ee
where ${\cal T}_I=2 {\cal T}(\rm{BH})\,{\rm Re}{\cal T}(\rm{DVCS})$.
The Bethe-Heitler contribution can be worked out without any approximation \ci{GKS} 
and is related to the electromagnetic form factors of the nucleon. The c.m.s. 
$\gamma^*p\to \gamma p$ amplitudes to leading-order of perturbation theory read  
\ba
\hspace*{-0.04\tw}{\cal M}_{\pm +,\pm +}(\gamma^*p\to\gamma p)\hspace*{-0.01\tw}&=
                     \hspace*{-0.01\tw} &\frac{e_0^2}{2}\sqrt{1-\xi^2}
               \Big[\langle H_{\rm eff}\rangle_\gamma \pm
                                 \langle \widetilde{H}_{\rm eff}\rangle_\gamma \Big]\,,\nn\\
\hspace*{-0.04\tw}{\cal M}_{\pm -,\pm +}(\gamma^*p\to\gamma p)\hspace*{-0.01\tw} &=&
   \hspace*{-0.01\tw}  -\frac{e_0^2}{2}
       \frac{\sqrt{t_0-t}}{2m} \Big[\langle E\rangle_\gamma \pm
                                 \langle \widetilde{E}\rangle_\gamma \Big]\,,
\ea
(compare with \req{eq:amplitudes}) and involve the convolutions
\ba
\langle K \rangle_\gamma &=& \int_{-1}^1 dx 
                      \Big[e_u^2 K^u+ e_d^2 K^d + e_s^2 K^s\Big]\nn\\
             &\times&          \,\Big[\frac1{\xi-x-i\varepsilon}  
             -\epsilon_k\,\frac1{\xi+x-i\varepsilon}\Big]
\label{eq:CFF-def}
\ea
where $\epsilon_k=+1$ for $K=H, E$ and $-1$ for $\widetilde{H}, \widetilde{E}$.  

The three terms in \req{eq:bh-interf-vcs} have the following harmonic
structure in $\phi$, the azimuthal angle of the outgoing photon with regard to
the leptonic plane (i=BH, DVCS):
\be
|{\cal T}_i|^2  \propto  L_i\sum_{n=0}^3 \left[ c_n^i \cos (n\phi ) + 
      s_n^i \sin (n\phi ) \right] 
\label{eq-cross-section} 
\ee
\noindent where $L_{\rm BH}=[-t P(\cos\phi)]^{-1}$ and $L_{\rm DVCS}=1$. An analogous 
Fourier series holds for the interference term. Although there are only harmonics 
up to the maximal order 3 in the sums, the additional $\cos\phi$ dependence from
the lepton propagators, included in $P(\cos{\phi})$, generates in principle an 
infinite series of harmonics for the BH and interference terms. A more detailed 
harmonic structure taking into account beam and target polarizations can be found 
in \ci{diehl-sapeta}. 

A comparison of this theoretical approach with experiment performed in \ci{kms}, 
reveals reasonable agreement with HERMES, H1 and ZEUS data and a less satisfactory 
description of the large-skewness, small $W$ JLab6 data. As discussed in Sect.\ 
\ref{sec:2} the application of the GPDs extracted from DVMP at JLab6 kinematics is 
problematic: It requires an extrapolation in $\xi$ and $t$, one has to be aware of 
possible soft-physics contributions (see e.g. Fig.\ \ref{fig:4}) as well as of large 
kinematical corrections. For example, in the relation \req{eq:xi-xb} there are 
additional terms proportional to $x_B$, e.g.\ $x_Bt/Q^2$ which are negligible in the 
small $\xi$,$-t$ region but not at JLab6 kinematics \ci{braun13}. 

It turns out that most of the $lp\to lp\gamma$ observables are under control of the 
best determined GPD $H$ (see Sect.\ \ref{sec:2} and Tab.\ \ref{tab:1}), only a few 
observables are sensitive to $E$ (e.g. particular modulations of $A_{UT}$) and 
$\widetilde H$ (e.g.\ the $\sin{\phi}$ modulation of $A_{UL}$). The GPD  $\widetilde E$ 
does not plays a role in DVCS in practice. 

\begin{figure}[t]
\centering
\includegraphics[width=0.33\tw]{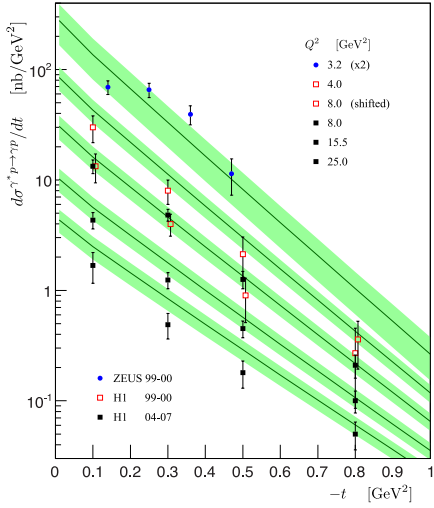}
\caption{The DVCS cross section for various values of $Q^2$ and $W$ ranging between
71 and $104\,\gev$. Data are taken from \ci{H1,zeus}. The results obtained in \ci{kms}
are shown as solid lines with error bands.}
\label{fig:7a}
\end{figure}

The DVCS cross section at HERA kinematics is shown in Fig.\ \ref{fig:7a}. Given that
this is a parameter-free calculation the agreement with experiment \ci{H1,zeus} is 
impressive. Similar results have been obtained in \ci{MM}. 

\begin{figure}[t]
\centering
\includegraphics[width=.27\tw]{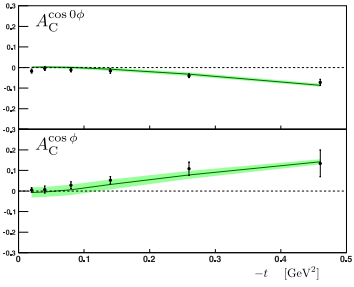}
\caption{The beam charge asymmetry versus $-t$ at $Q^2\simeq 2.51\,\gev^2$, 
$x_B\simeq 0.097$. Data are taken from \ci{hermes-bsa}. For further notations see 
Fig.\ \ref{fig:7a}.}
\label{fig:8}
\end{figure}

The $lp\to lp\gamma$ cross section on an unpolarized target for a given
beam charge, $e_l$, and beam helicity, $h_l/2$, can be decomposed as
\ba
d\sigma^{h_l,e_l}(\phi)&=& d\sigma_{\rm UU}(\phi)\Big[ 1+ h_l A_{\rm LU, DVCS}(\phi) \nn\\
                     &+& e_lh_lA_{\rm LU,I}(\phi) + e_lA_C[\phi)\Big]
\label{eq:decomposition}
\ea
where only the $\phi$-dependence of the observables is made explicit. If a
longitudinally polarized beam of both lepton charges is available the asymmetries  
in \req{eq:decomposition} can be isolated. Two modulations of the beam charge asymmetry, 
$A_C$, are shown in Fig.\ \ref{fig:8}. They are under control of the BH-DVCS interference 
and depend mostly on ${\rm Re}\langle H \rangle_\gamma$. The agreement of the handbag 
results with the HERMES data \ci{hermes-bsa} demonstrates that the convolution of $H$ has 
the right magnitude. 

\begin{figure}[t]
\centering
\includegraphics[width=.25\tw, bb=515 1 1022 494,clip=true]{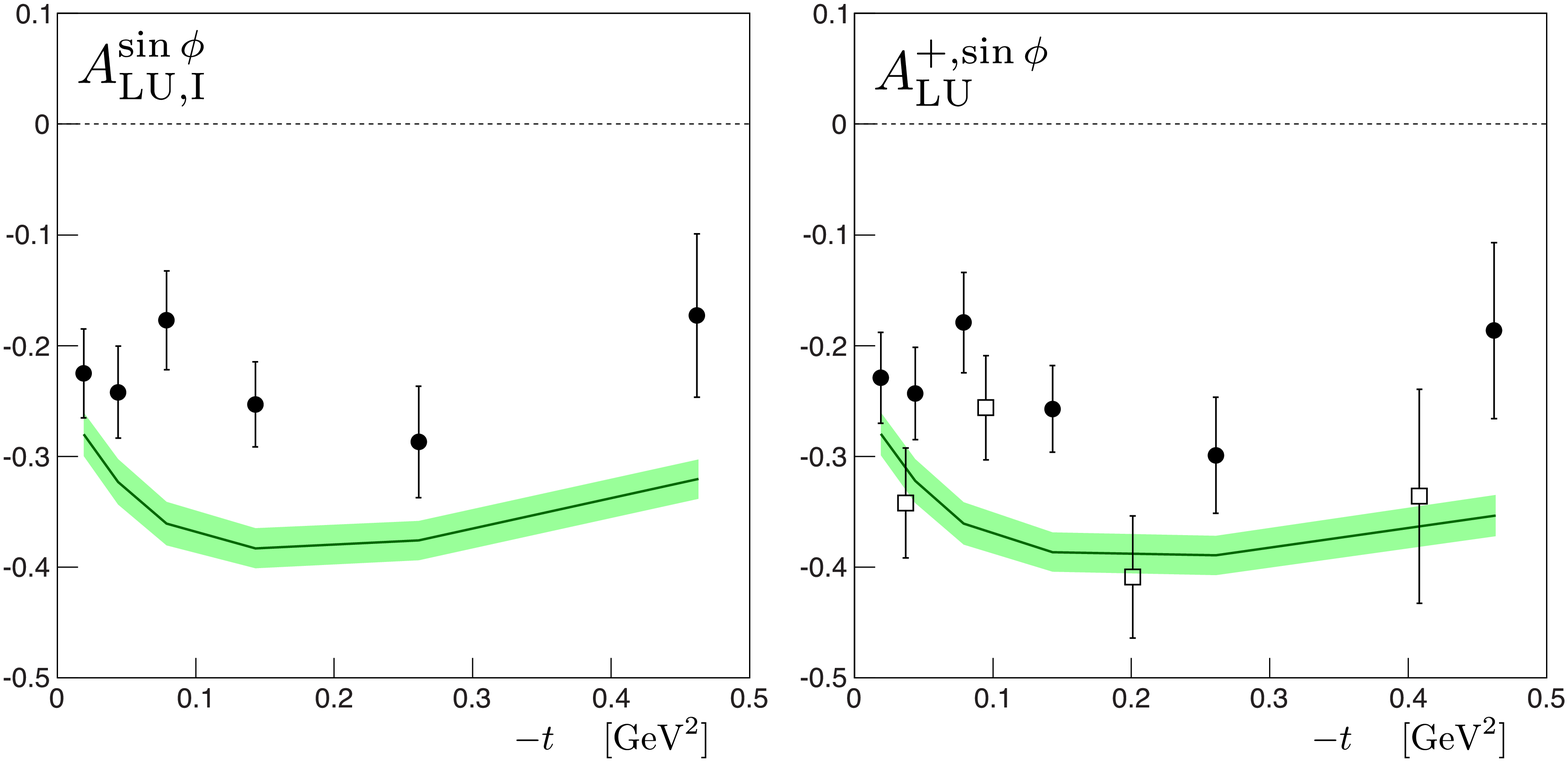}
\caption{The $\sin{\phi}$ modulation of $A_{LU}^+$ versus $-t$ at
$Q^2\simeq 2.5\,\gev^2$, $x_B\simeq 0.09$. Data are taken from \ci{hermes-bsa} (solid 
circles) and \ci{hermes-recoil} (open squares). For further notations see Fig.\ 
\ref{fig:7a}.}
\label{fig:9}
\end{figure}

The HERMES collaboration has measured the $\sin \phi$ modulation of the beam spin 
asymmetry using a recoil detector \ci{hermes-recoil} which allows for a detection of all 
three final-state particles. The resonant background is therefore severely reduced and in 
so far the recoil data are closer to the exclusive process $lp\to lp\gamma$ to which the 
theory applies. Since the recoil data are available only for a positron beam the 
observables defined in \req{eq:decomposition} cannot be isolated. In fact, the combination
\be
A_{\rm LU}^{+\sin \phi} \simeq \frac{A_{\rm LU,I}^{\sin \phi}}{1+A_C^{\cos(0 \phi)}}
\label{eq:recoil}
\ee
is measured by the HERMES collaboration. In \req{eq:recoil} a contribution from 
$A_{\rm LU,DVCS}$ is neglected which is exactly zero at twist-2 accuracy in agreement with 
experiment \ci{hermes-bsa}. Since $A_C^{\sin{(0\phi)}}$ is so small $A_{\rm LU}^{+,\sin{\phi}}$
and $A_{\rm LU,I}^{\sin{\phi}}$ don't differ much. As can be seen from Fig.\ \ref{fig:9} 
the handbag results for $A_{\rm LU}^{+\sin \phi}$ agree quite well with the recoil data 
while the agreement with the non-recoil data \ci{hermes-bsa} is surprisingly bad. 
Recoil data for other observables would be welcome. 

Results on the transverse target spin asymmetries will be discussed in the next section.

\section{The GPD $E$}
\label{sec:5}
\begin{figure*}[t]
\centering
\includegraphics[width=0.7\tw,bb=43 49 594 301,clip=true]{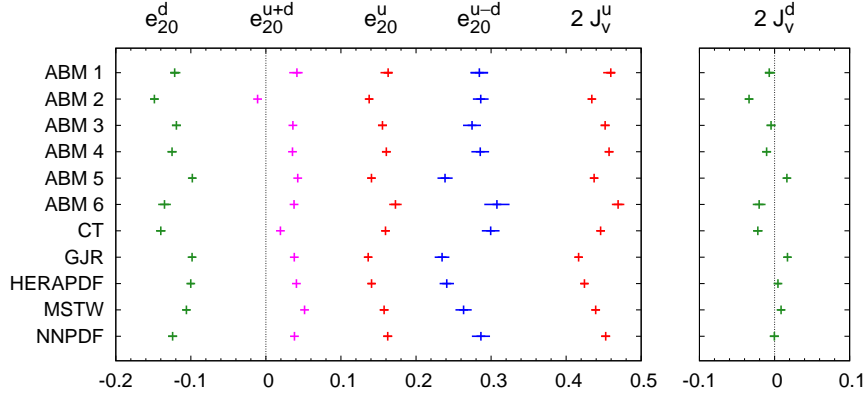}
\caption{Results for the second moments of $E$ and the angular momenta of the valence quarks 
at a scale of $2\,\gev$. Shown is the default fit ABM1 and variations with regard to the 
strangeness form factors, other data interpolations and different sets of PDFs \ci{DK13}.} 
\label{fig:10}
\end{figure*}

Next let me discuss the GPD $E$ in some detail. The analysis of the nucleon form 
factors performed in \ci{DFJK4} and updated in \ci{DK13}, provides the zero-skewness 
GPDs for valence quarks which can be used as input to the DD representation 
\req{eq:int-rep}. The basis of this analysis are the sum rules for the flavor form 
factors 
\ba
F_1^q(t)&=&\int_0^1 d\rho H_v^q(\rho,\xi=0,t)\,,\nn\\
F_2^q(t)&=&\int_0^1 d\rho E_v^q(\rho,\xi=0,t)
\ea
where the valence quark GPDs are defined by
\be
K_v^q(\rho,\xi=0,t)\=K^q(\rho,\xi=0,t)+K^q(-\rho,\xi=0,t)\,.
\ee
The Dirac ($i=1$) and Pauli ($i=2$) form factors for the proton and the neutron are 
decomposed in the flavor form factors as
\be
F_i^p= e_uF_i^u+ e_dF_i^d+e_s F_i^s\,, \quad F_i^n= e_uF_i^d+ e_dF_i^u+e_s F_i^s\,.
\ee
Estimates say that the strangeness form factors can be neglected, their contributions
are at most of the size of the errors on $F_i^u$ and $F_i^d$. The 
valence-quark GPDs are parametrized as in \req{eq:zero-skewness} and 
\req{eq:profile-dfjk4}. The forward limits of $E$ are parametrized as in \req{eq:forward};
in \ci{DK13} an additional factor $(1+\gamma_q\sqrt{\rho})$ is used with $\gamma_u=4$ and 
$\gamma_d=0$. This factor improves $\chi^2$ slightly. The GPD $E$ is constrained by
\be
\int_0^1 d\rho e_v^q(\rho)\=\kappa_q
\ee
where $\kappa_q$ is the contribution of quarks of flavor $q$ to the anomalous magnetic
moments of the nucleon ($\kappa_u=1.67$, $\kappa_d=-2.03$). In Sect.\ \ref{sec:2} it has 
already been remarked that the 2004 analysis \ci{DFJK4} is based on the CTEQ6M PDFs 
\ci{cteq6} while the recent update \ci{DK13} uses the ABM11 PDFs \ci{abm11} for the 
default fit. Since in 2004 data on the neutron form factors were only available for 
$-t\lsim 2\,\gev^2$ the parameters of the zero-skewness GPD $E$ were not well fixed; a 
particularly wide range of values were allowed for the powers $\beta^u_e$ and $\beta^d_e$. 
In the reanalysis \ci{DK13} use is made of the new data on the neutron form factors  
and the ratio of electric and magnetic proton form factors which extend to much larger 
values of $-t$ than before \ci{lachniet,riordan,gayou02,puckett}. Because of the 
$\rho -t$ correlation discussed above, the powers $\beta^q_e$ are better determined now 
($\beta_u=4.65$, $\beta_d=5.25$, $\alpha_u(0)=\alpha_d(0)=0.603$). At small $-t$ the new 
results for the valence-quark GPDs are similar to the 2004 version. As an example of the 
results derived in \ci{DK13} the second moments of $E$ for valence quarks at $\xi=t=0$ 
are displayed in Fig.\ \ref{fig:10}.  
\begin{figure*}[th]
\centering
\includegraphics[width=0.34\tw]{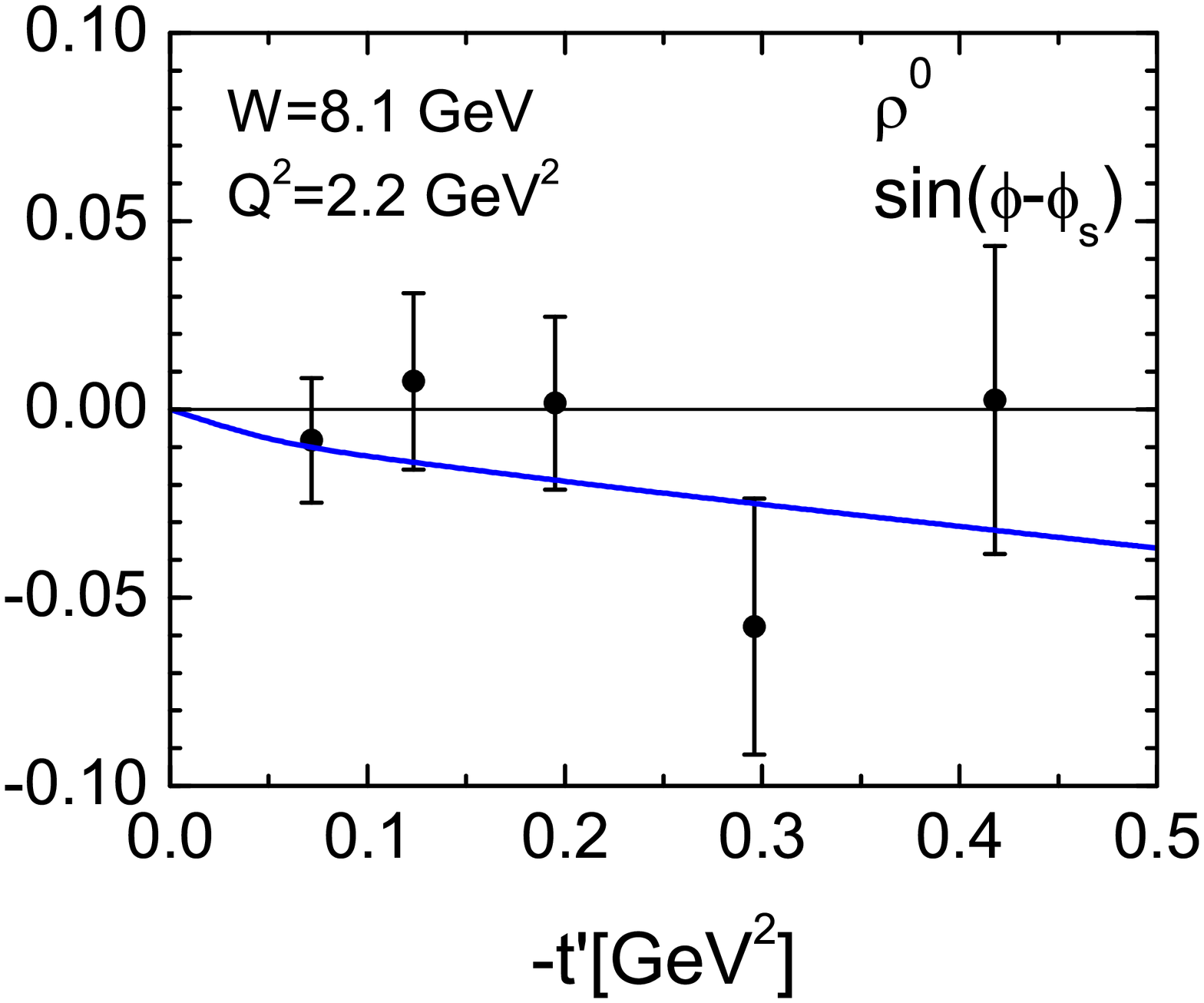} \hspace*{0.15\tw}
\includegraphics[width=0.28\tw,bb=1 1 253 248,clip=true]{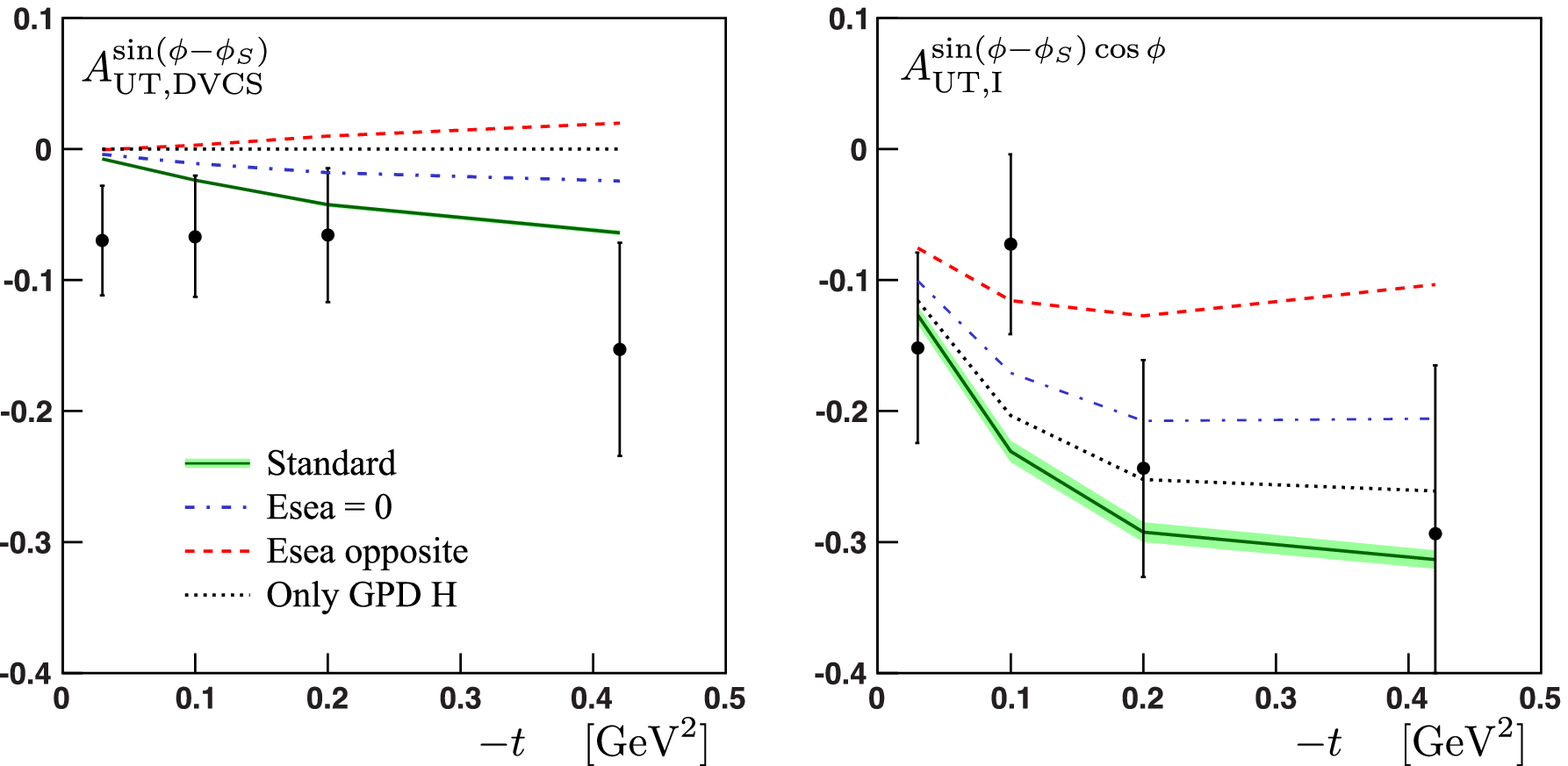}
\caption{Left (right):The $\sin{(\phi -\phi_s)}$ modulation of $A_{\rm UT}$ for $\rho^0$ 
production (DVCS). Data are taken from \ci{compass-aut-rho} (\ci{hermes-aut-dvcs}), 
the theoretical results from \ci{GK7} (\ci{kms}).} 
\label{fig:11}
\end{figure*}

For the following discussion it is convenient to  change a little bit the notation.
In analogy to the PDFs we define
\ba
\hspace*{-0.06\tw} e^q(\rho)\hspace*{-0.01\tw} &=&\hspace*{-0.01\tw}   E_q(\rho,\xi=t=0)\,, 
\; e^{\bar{q}}(\rho)\=E_q(-\rho,\xi=t=0)\,, \nn\\
\hspace*{-0.05\tw} \rho e^g(\rho)\hspace*{-0.01\tw} &=&\hspace*{-0.01\tw}  E_g(\rho,\xi=t=0)\,.  
\ea
The Mellin moments are defined as
\be
e^q_{20}\=\int_0^1d\rho \rho^{n-1} e^q(\rho)\,, \quad e^g_{20}\=\int_0^1d\rho \rho^{n-1} e^g(\rho)\,,
\ee
and analogously for the PDFs.

Not much is known about $E^g$ and $E^{\rm sea}$. 
There is only a sum rule for the second moments of $E$ \ci{teryaev99} at 
$t=\xi=0$ 
\be
e_{20}^g\=-\sum e_{20}^{q_v} -2\sum e_{20}^{\bar{q}}\,.
\label{eq:SumRuleE}
\ee 
It turns out that the valence contribution to the sum rule is very small. In fact,
$e_{20}^{u_v}+ e_{20}^{d_v}=0.041^{+0.011}_{-0.053}$ for the default fit, see Tab.\ 18 in 
\ci{DK13}. Hence, the second moments of the gluon and sea-quark GPD $E$ cancel to a
large extent. Since the parametrization \req{eq:forward} for the forward  limit 
of $E$ does not have nodes except at the end-points this property approximately 
holds for other moments as well and even for the convolution \req{eq:convolution-mesons}.

A further information about $E$ for strange quarks comes from a positivity bound 
for its Fourier transform \ci{burkardt03}: 
\be
\frac{b_\perp^2}{m^2}\left(\frac{\partial e_s(\rho,b_\perp)}{\partial b_\perp^2}\right) \leq 
     s^2(\rho,b_\perp) - \Delta s^2(\rho,b_\perp)
\label{eq:positivity}
\ee
where $s$, $\Delta s$ and $e_s$ are the Fourier transforms of the zero-skewness GPDs $H^s$, 
$\widetilde{H}^s$ and $E^s$, respectively. As shown in \ci{diehl-kugler,GK4} this bound 
forbids a large strange quark contribution and, assuming a flavor-symmetric sea, a large 
gluon contribution too. The bound on $e^s(\rho,b_\perp)$ is saturated for $N_e^s=\pm 0.155$ 
($\beta_e^s=7$ and the same $\delta_s$ as for $H^s$ in \req{eq:forward}) \ci{GK4}.
The normalization of $e^g$ can subsequently be fixed from the sum rule \req{eq:SumRuleE} 
($\beta^g_e=6$ and the same $\delta_g$ as for $H^g$). These results are 
inserted in \req{eq:int-rep} in order to obtain estimates of $E^{\rm sea}$ and $E^g$.
Diehl in \ci{eic} has studied the GPD $E$ along the same lines as discussed here.

The GPD $E$ is probed by the transverse target spin asymmetry~\footnote{
$\phi_s$ is the orientation of the target spin vector with respect to the lepton plane.} 
\be 
 A_{UT}^{\sin{(\phi-\phi_s)}} \sim {\rm Im}\Big[\langle E \rangle^*\langle H\rangle \Big]\,,
\ee
for given $H$ \ci{GK2}. 
The described parametrization of $E$ is consistent with the data on $\rho^0$ 
production from HERMES \ci{hermes-aut-rho} and COMPASS \ci{compass-aut-rho} 
(see Fig.\ \ref{fig:11}). However, only $E$ for valence quarks matters for $A_{UT}$ 
since the sea and gluon contribution to $E$ cancel to a large extent. Fortunately 
the analysis of DVCS data \ci{kms} provides additional although not very precise 
information on $E^{\rm sea}$. To leading-order of pQCD there is no gluon contribution
in DVCS and therefore $E^{\rm sea}$ becomes visible. The data on the $\sin{(\phi-\phi_s)}$
modulation of the transverse target spin asymmetry for DVCS measured by the HERMES 
collaboration \ci{hermes-aut-dvcs} are shown on the right hand side of Fig.\ 
\ref{fig:11} and compared to the results obtained in \ci{kms}. Despite the large 
experimental errors a negative $E^{\rm sea}$ seems to be favored. Independent 
information on $E^g$ would be of interest. This may be obtained from a measurement of 
the transverse target polarization in $J/\Psi$ photoproduction \ci{koempel}. 

\section{Ji's sum rule}
\label{sec:6}
\begin{figure}[t]
\centering
\includegraphics[width=.38\tw]{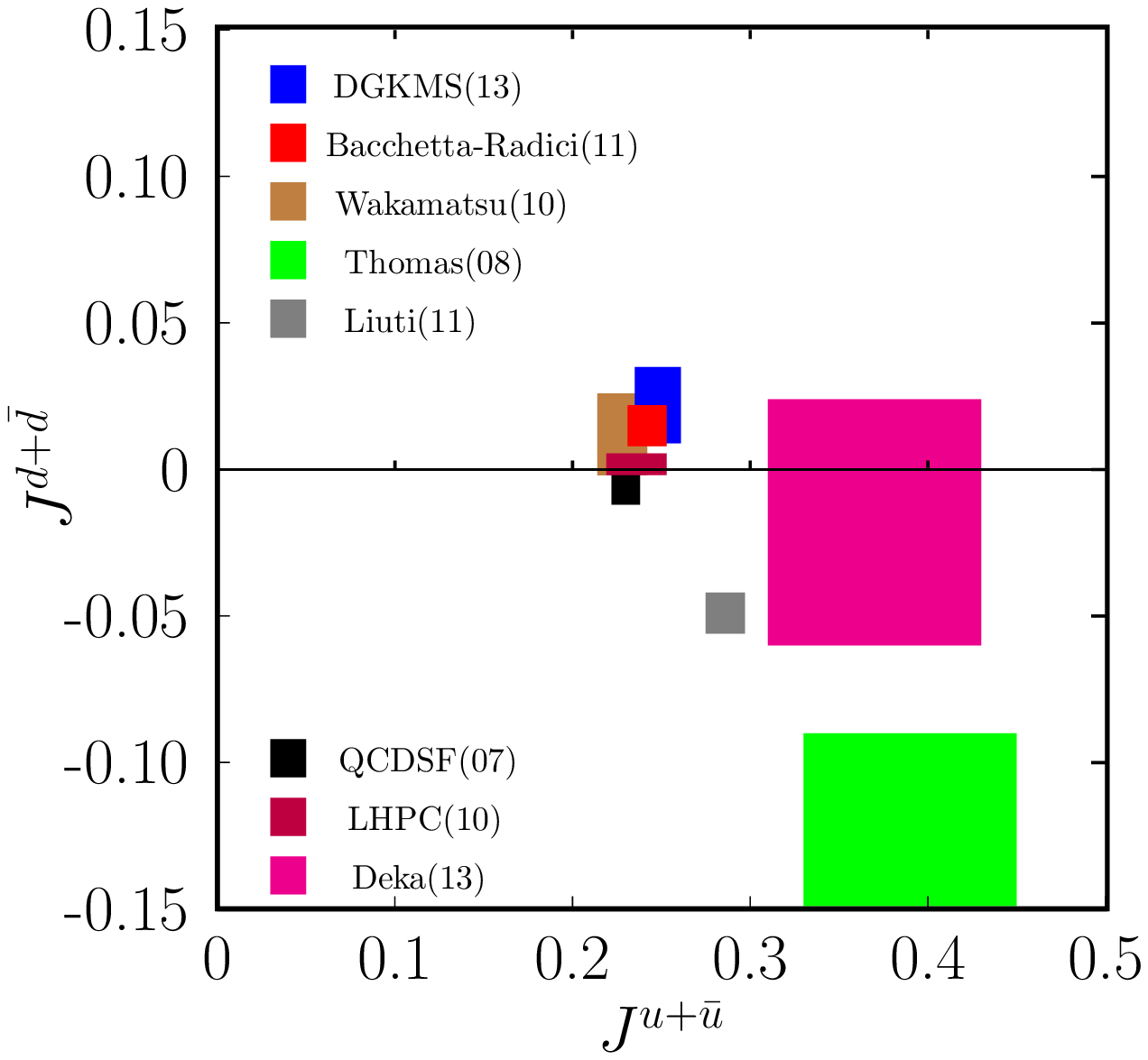}
\caption{Results on the angular momenta for $u$ and $d$ quarks. Data are from
the combined analyses \ci{DK13,GK7,kms} and from \ci{bacchetta,wakamatsu,thomas,
liuti,qcdsf07,lhpc,deka}.}  
\label{fig:13}
\end{figure}

The knowledge of the GPDs allow for an evaluation of the angular momenta the
partons inside the proton carry. At $\xi=t=0$ the angular momenta~\footnote{
For a proton that moves along the 3-direction, $J$ is the expectation value of the 
3-component of the parton angular momentum operator.}
are given by the sum of the second moments of the PDFs and the $\xi=t=0$ limits of 
$E$ \ci{ji96} ($q=u,d,s,\bar{u},\bar{d},\bar{s}$)
\be
J^q\=\frac12\Big[q^q_{20}+e_{20}^q\Big]\,, \qquad J^g\=\frac12\Big[g_{20}+e_{20}^g\Big]
\ee
The analysis of the nucleon form factors \ci{DK13} provides 
\be
J^{u}_v\=0.230^{+0.009}_{-0.024}\,, \qquad J^{d}_v\=-0.004^{+0.011}_{-0.017}\,
\label{eq:Jval}
\ee
for the valence quarks. For an evaluation of $J$ for all quarks and the gluon the moments 
from the ABM11 PDFs are used \ci{abm11}, the results \req{eq:Jval} on $J$ for valence 
quarks and the estimate of $e^s_{20}$ from the positivity bound \req{eq:positivity}
and the analysis of $A_{UT}$ for DVCS \ci{kms}: $e^s_{20}=0.0\ldots -0.026$. Assuming
a flavor symmetric sea for $E$ one subsequently fixes the $2^{\rm nd}$ moment of $E_g$  
from the sum rule \req{eq:SumRuleE} ($e_{20}^g=-0.041\ldots 0.115$). Combing all information
on the second moments, one obtains 
at the scale $2\,\gev$ 
\ba
\hspace*{-0.05\tw}J^{u+\bar{u}}&=& 0.261\ldots 0.235\,, \quad 
                     J^{s+\bar{s}}=0.017\ldots -0.009\,, \nn\\
\hspace*{-0.05\tw} J^{d+\bar{d}}&=& 0.035\ldots 0.009\,,  \quad 
                     J^{\,g}\;\,= 0.187\ldots 0.265\,. 
\label{eq:J-results}
\ea
The values of the left-hand side are evaluated from $e^s_{20}=0.0$, those on the right-hand
side from $e^s_{20}=-0.026$. Thus, the badly known $E^s$ determines the uncertainties of 
the angular momenta at present. Data on $A_{UT}$ for DVCS with smaller errors than obtained 
by HERMES \ci{hermes-aut-dvcs} would reduce the errors on the angular momenta. The large 
value of $J^g$ is no surprise. The value of $g_{20}$ represents the familiar result that 
about $40\%$ of the proton's momentum is carried by the gluons. Since $|e^g_{20}|$ seems to 
be much smaller than $g_{20}$ this result is not changed much. The angular momenta in 
\req{eq:J-results} sum to 1/2, the spin of the nucleon, because the sum rule 
\req{eq:SumRuleE} is used in the analysis 
and the PDFs respect the momentum sum rule of DIS. It is to be stressed that the results 
are obtained from a combination of inclusive (the PDFs) and exclusive data (form factors, 
DVMP, DVCS). This differs from attempts to understand the nucleon spin only from DIS. A 
comparison of different results on the angular momenta is made in Fig.\ \ref{fig:13}. 
There are also experimental results on the angular momenta  extracted from DVCS data: 
$J^{d+\bar{d}} + J^{u+\bar{u}}/5=0.18\pm 0.14$ by \ci{clas-mazouz} and 
$J^{d+\bar{d}}/2.9 + J^{u+\bar{u}}=0.42\pm 0.21\pm 0.06$ by HERMES \ci{hermes-ye}. These 
results are strongly model-dependent. Among other things they rely on the assumption of 
proportionality between $e^{q_v}$ and $q_v$ which is in conflict with the form factor 
analysis \ci{DK13} and with perturbative QCD arguments \ci{yuan04}. In a recent lattice 
QCD study \ci{bali} the lowest moment of the isoscalar quark distribution, $u_{20}-d_{20}$, 
has been calculated for pion masses ranging from 157 till $500\,\mev$. Substantial 
contributions from excited states to the nucleon structure have been found. After their 
subtraction the moment $u_{20}-d_{20}$ is much smaller ($\simeq 0.2$) than obtained in 
other lattice QCD studies ($\simeq 0.26$ as for instance in \ci{deka}) but still larger
than found in PDF analyses ($\simeq 0.16$), e.g.\ \ci{abm11}. Thus, one has to be cautious
in applying lattice QCD results, there may still be substantial uncertainties.  

The orbital angular momenta are obtained
from $J$ by subtracting the first moment of the polarized PDFs
\be
L^q\=\frac12 \Big[q_{20}+e^q_{20}-\Delta q_{10}\Big]\,.
\ee
Taking the polarized PDFs from \ci{dssv}, one obtains for the valence quarks the values
\be
L^u_v\=-0.141^{+0.025}_{-0.033}\,, \qquad L^d_v\=0.114^{+0.034}_{-0.035}\,,
\ee
and for quarks generally
\ba
\hspace*{-0.05\tw}L^{u+\bar{u}}\hspace*{-0.02\tw}&=&\hspace*{-0.02\tw}-0.146\ldots -0.172\,, 
\quad 
L^{d+\bar{d}}\= 0.263\ldots 0.237\,,\nn\\
\hspace*{-0.05\tw} L^{s+\bar{s}}\hspace*{-0.02\tw} &=&\hspace*{-0.02\tw} \phantom{-}
0.073\ldots \phantom{-} 0.047\,.
\label{eq:L-results}
\ea
A reliable decomposition of $J^g$ in spin and orbital angular momentum is not possible
at present \ci{dssv14}.


\section{Summary}
\label{sec:7}
I have summarized the recent progress in the analysis of hard exclusive
leptoproduction of mesons ($\rho^0$, $\phi$, $\omega$, $\pi^+$) and photons at small 
skewness and small $-t$ within the handbag approach. A set of GPDs has been extracted 
which is constructed from double distributions with parameters adjusted to meson 
leptoproduction data ($\rho^0$, $\phi$ and $\pi^+$) and nucleon form factors. This set 
of GPDs allows for a parameter-free calculation of DVCS and has also been   
used in an analysis of $\omega$ leptoproduction. Very good agreement is found with the 
SDMEs measured by the HERMES collaboration \ci{hermes-omega}. Interesting predictions 
have also been given for $\pi^0$ and $\eta$ production where the transversity GPDs
seem to dominate. Lack of small $\xi$, small $-t$ data prevents a verification of these 
predictions at present. Large skewness data from JLab6, however, do agree with the 
predictions in tendency.  

There are many observations that the experimental data do not agree with the naive
asymptotic results obtained in collinear approximation (leading-twist accuracy) and 
leading-order of perturbation theory. In particular at JLab6 kinematics, characterized 
by large skewness and small $W$, the application of the handbag approach is problematic -  
meson leptoproduction in this kinematical region is not understood as yet. One has to 
be aware of eventual soft-physics corrections in some of the reactions. In any case an 
application of the GPDs derived in \ci{GK2,GK3,GK5} to the region of JLab6 kinematics 
requires their extrapolation to large $\xi$ and large $-t$.

\textit{Acknowledgements}: The authors thanks Umberto D'Alesio and Francesco Murgia 
for the kind invitation to the interesting and excellent organized workshop on Transversity
in Chia (Sardinia).

\end{document}